\documentclass[traditabstract]{aa}

\usepackage{epsfig}
\usepackage{graphics}
\usepackage[latin1]{inputenc}
\usepackage{natbib}
\usepackage{ulem} 

\def\zpho{{z_{\rm pho}}}
\def\zgen{{z_{\rm gen}}}
\def\zspe{{z_{\rm spe}}}
\def\zgal{{z_{\rm gal}}}

\begin{document}
\title {Photometric redshifts for Type Ia supernovae in 
the Supernova Legacy Survey}
\author{N. Palanque-Delabrouille\inst{1} \and V. Ruhlmann-Kleider\inst{1} \and S. Pascal\inst{1} \and
J. Rich\inst{1}\and J. Guy\inst{2} \and G. Bazin\inst{1}, 
P. Astier\inst{2}, C. Balland\inst{2,3},
S. Basa\inst{4},
R. G. Carlberg\inst{6}, A. Conley\inst{10}, D. Fouchez\inst{7}, 
D. Hardin\inst{2}, I. M. Hook\inst{9,11}, D. A. Howell\inst{12,13},
R. Pain\inst{2}, K. Perrett\inst{6},
C. J. Pritchet\inst{8}, N.  Regnault\inst{2} \and
M. Sullivan\inst{9}
}
\institute{CEA, Centre de Saclay, Irfu/SPP,  F-91191 Gif-sur-Yvette, France 
\and  LPNHE, Université Pierre et Marie Curie, Université Paris diderot, CNRS-IN2P3, 4 place Jussieu, 75252 Paris Cedex 05, France
\and University Paris 11, Orsay, F-91405, France
\and LAM, CNRS, BP8, P\^{o}le de l'\'etoile, Site de Ch\^{a}teau-Gombert,
38 rue Fr\'ed\'eric Joliot-Curie, F-13388 Marseille Cedex 13, France
\and LUTH, UMR 8102 CNRS, Observatoire de Paris, Section de Meudon, F-92195 Meudon Cedex, France
\and Deparment of Astronomy and Astrophysics, University of Toronto, 50 St. George Street, Toronto, ON M5S 3H8, Canada
\and CPPM, CNRS-Luminy, Case 907, F-13288 Marseille Cedex 9, France
\and Department of Physics and Astronomy, University of Victoria, PO Box 3055, Victoria, BC V8W 3P6, Canada
\and Department of Physics (Astrophysics ),University of Oxford, Denys Wilkinson Building, Keble Road, Oxford OX1 3RH, UK
\and Center for Astrophysics and Space Astronomy, University of Colorado, Boulder, CO 80309-0389, USA
\and INAF - Osservatorio Astronomico di Roma, via Frascati 33, 00040 Monteporzio (RM), Italy 
\and Las Cumbres Observatory Global Telescope Network, 6740 Cortona Dr.,
Suite 102, Goleta, CA 93117
\and Department of Physics, University of California, Santa Barbara, Broida
Hall, Mail Code 9530, Santa Barbara, CA 93106-9530
}
\date{Received xx; accepted February 15, 2010}
\authorrunning{N. Palanque-Delabrouille et al.}
\titlerunning{Photometric redshifts for SNIa in SNLS}
\abstract{We present a method using the SALT2 light curve fitter to determine the redshift of Type Ia supernovae in the Supernova Legacy Survey (SNLS) based on their photometry in $g'$, $r'$, $i'$ and $z'$. On 289 supernovae of the first three years of SNLS data, we obtain a precision  $\sigma_{\Delta z/(1+z)} = 0.022$ on average up to a redshift of 1.0, with a higher precision of $ 0.016$ for $z<0.45$ and a lower one of $ 0.025$ for $z>0.45$. The rate of events with $|\Delta z|/(1+z)>0.15$ (catastrophic errors) is 1.4\%, and reduces to 0.4\% when restricting the test sample of spectroscopically confirmed Type Ia supernovae. Both the precision and the rate of catastrophic errors are better than what can be currently obtained using host galaxy  photometric redshifts. Photometric redshifts of this precision may be useful for future experiments which aim to discover up to millions of supernovae Ia but without spectroscopy for most of them.
 }
\maketitle

\section{Introduction}

Ten years after the discovery of the accelerating expansion of the Universe, Type Ia supernovae (SNe Ia) are still among the most accurate probes of dark energy. To study systematic effects more precisely, future experiments aim at an increase of several orders of magnitudes in the number of detected supernovae.
Pan-STARRS Medium Deep Survey (MDS) and the Dark Energy Survey will discover thousands of supernovae up to a redshift of 1, though they will only be able to observe spectroscopically a fraction of them. The Large Synoptic Survey Telescope will discover roughly 250,000 Type Ia SNe per year in the same redshift range, but again with only partial spectroscopic coverage (see the white paper of \cite{bib:howell09} for a review of future surveys).  
While current surveys as the Supernova  Legacy Survey \citep{bib:astier} are already limited by the time available for spectroscopy, most future surveys will indeed have to count on an alternative to select the SNe Ia and  determine their redshift from light curves only. 

Studies solely based on photometry present the advantage of leading to larger sets of supernovae, that can be used to compare properties of SNe Ia in various subsamples, sorted for instance by redshift, SN color, or host morphology. The possibility to use solely photometric redshifts to derive cosmological constraints from SNe Ia, however, remains to be shown. 

This paper addresses the question of obtaining photometric redshifts for SNe Ia. It presents a method that can be applied to ongoing as well as to future projects. The combination of a photometric selection of supernovae in the SNLS data, which doubles the number of supernovae compared to the real-time selection, with the photometric redshifts derived according to the method presented in this paper, will be the subject of a forthcoming study. 

Previous works have already tackled the topic of SN Ia photometric redshifts. An empirical photometric redshift estimate for SNe Ia  was proposed by~\cite{bib:wang} who obtained a rms dispersion of 0.050 for 20 supernovae from the SNLS first year data not contained in the training set, and 0.031 for 20 supernovae used in the training. A method was also developed in SNLS for the needs of the real-time selection of SN Ia candidates  to send for spectroscopy~\citep{bib:sullivan06}. It used the rising part of the light curves and a cosmology prior to both select tentative SN Ia candidates and derive an estimate of their redshift. Fits to the pre-maximum light data reached a precision $\sigma_{|\Delta z|/(1+z)} = 0.046$ while fits to the entire light curve a precision of 0.037.

The method presented in this paper aims at an improved precision on the photometric redshift, while remaining independent of the underlying cosmology.
 The study was calibrated with 
simulated SN Ia light curves, whose characteristics are given in \S 2. The method is presented in \S 3. To test its performance, it was applied to a large sample of SNe Ia from the first three years of SNLS data, presented in \S 2. The results obtained on this test sample are given in \S 4.
A discussion on the method and on the impact of the underlying cosmology is presented in \S 5. The possible redsfhift biases and the precision of the method for various sub-classes of SNe Ia is discussed in \S 6. Conclusions are then given in \S 7. 

\section{SNLS data and simulation description}

 The SNLS data cover the redshift range between 0 and 1.2. They consist of four-band light curves obtained in a rolling search mode, leading to roughly 30 points per observing season in $r'$ and $i'$, 20 points in $g'$ and 15 points in $z'$. The test sample used to quantify the perfomance of the method consists of a total of 289 SNe Ia passing color and stretch cuts ($|c|<0.4$, $|x_1| < 4$) to avoid events with extreme values compared to typical SNe Ia. It originates from the union of two sources:  
241 are real-time selected events confirmed as ``Ia" or ``Ia?" by spectroscopy \citep{bib:howell, bib:bronder, bib:ellis, bib:balland}, and the additional 48 are issued from a SN Ia photometric selection by  Bazin et al. (in prep.) based on the host galaxy photometric redshift. The latter SNe Ia have a host spectroscopic redshift (thus allowing their use in the test sample) but no type determination because of insufficient S/N for the supernova spectrum. The contamination of the photometric selection of SNe Ia is estimated to be 7\%, implying a possible contamination 3.4 events out of 48. Our total sample of 289 events is thus expected to be contaminated at the 1\% level. 

\begin{figure}[h]
\includegraphics[width=\columnwidth]{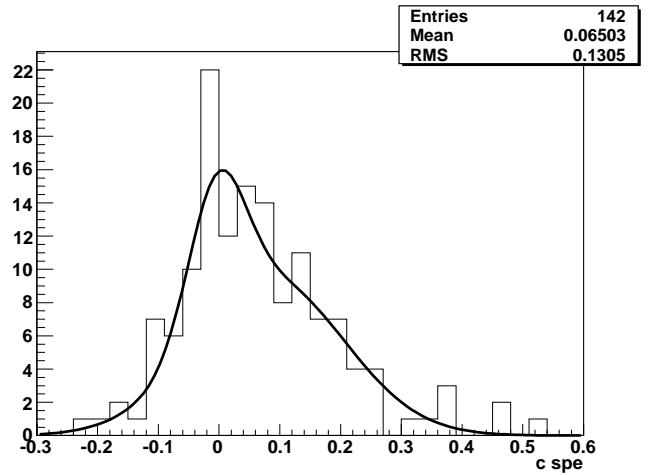}
\caption{
Color distribution of SNe Ia with redshift $z<0.7$ from the first three years of SNLS data. The curve is a fit to the data with a double Gaussian as explained in the text.
}
\label{fig:color}
\end{figure}

Synthetic SN Ia light curves were produced with SALT2 \citep{bib:guy2007} assuming a flat Universe with $\Omega_M = 0.26$. The light curves are characterized by five parameters: color $c$, stretch $x_1$, absolute normalization, date  of maximum flux in the rest frame $B$ band and redshift. Redshifts were simulated in the range from 0 to 1.2 assuming a constant volumetric SN Ia rate.  The date of maximum was drawn uniformly within the dates of observation in SNLS. Values of stretch were generated according to a Gaussian distribution (centered on $m= 0$ and with standard deviation $\sigma= 1$), and those of color according to a double Gaussian distribution ($m_1 = 0.00$, $\sigma_1=0.05$, $m_2 = 0.08$, $\sigma_2=0.12$, and relative amplitudes $A_1/A_2 = 0.91$, see figure~\ref{fig:color}). These distributions best reproduce our selected data sample at redshift $z<0.7$ where Malmquist bias is negligible (cf. \cite{bib:astier} fig. 13).

Model light curves in the four bands of SNLS ($g'$, $r'$, $i'$ and $z'$) were computed. Instrumental effects were taken into account following the distributions of flux uncertainties and dispersions observed in data, as well as correlations between flux uncertainties and flux values. A total of 20000 SNe Ia were generated. A preselection filter was applied to these events to remove those where the light curve was either too faint or too poorly sampled to be distinguished from pure noise, leaving 11106 SNe Ia.  
More details on the simulation and the preselection can be found in Bazin et al. (in prep.).

\section{Redshift fit}\label{sec:procedure}

The light curve fitter used for this analysis is based on the public version of SALT2 with additions described below. The $\chi^2$ minimization is done with the Minuit package.\footnote{http://wwwasdoc.web.cern.ch/wwwasdoc/minuit/minmain.html}

\begin{figure}[htb]
 \vskip -.3cm \includegraphics[width=\columnwidth]{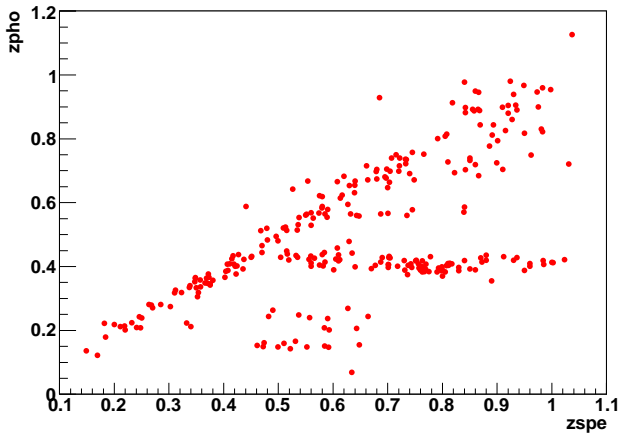}
\caption{
Photometric vs. spectroscopic redshift  for the data sample in the absence of initialization or prior on redshift.
}
\label{fig:zpho_v0}
\end{figure}

 By default, the fitting procedure deals with redshift as with the other free parameters, using initial values of 0, 0 and $\sim 0.1$ respectively for the parameters describing the color $c$, the stretch $x_1$ and the redshift of the supernova. The fit consists of a $\chi^2$ minimization of the multi-band light-curves according to an empirical modelling of the SN Ia luminosity variations as a function of phase, wavelength, $x_1$, and $c$. Figure~\ref{fig:zpho_v0} shows the distribution of photometric redshift ($\zpho$) vs. spectroscopic redshift ($z_{\rm spe}$) obtained using the default SALT2 to fit $\zpho$. 
Despite the use of priors on color and stretch defined by the statistical distribution of these two parameters (see section~\ref{sec:priors}), 43\% of the events had $|\zpho - \zspe|/(1+\zspe) \equiv |\Delta z|/(1+z) > 0.07$, and some redshifts, such as $z_{\rm pho} \sim 0.4$, were clearly favored over others. This shows that an initial step is required in order to start the $\chi^2$ minimization with  reasonable values of the parameters for each individual supernova (not surprisingly, as the fit is non-linear). The procedure to initialize the fit is described in section~\ref{sec:proc}. 

The best precision one can achieve with SNLS data can be estimated by initializing the fit at the actual value of the redshift, as was done in~\cite{bib:guy2007}. This led to a precision (cf. definition in section~\ref{sec:accuracy}) of 0.018  and only 3\% of events with $|\Delta| z/(1+z) > 0.07$. The main goal of the study presented in this paper was to approach this ideal situation by determining, prior to the full fit, the most accurate initial values not only for redshift but also for color and stretch.

Since different combinations of the five parameters that describe the four-band light curves of a SN Ia could yield similar $\chi^2$ (although giving extreme values of color or stretch), it was essential to guide the fitter with priors in order to converge to the most accurate values (cf. section~\ref{sec:priors}).

\subsection{Priors and bounds}\label{sec:priors}

Several of the parameters $p_j$ were constrained by Gaussian priors (mean $m_j$ and variance $v_j$) which were included in a generalized $\chi^2$ as follows: 
\begin{equation}
\chi^2 = \sum_i \frac{[f_{\rm data}(t_i) - f_{\rm model}(t_i)]^2}{\sigma(t_i)^2} + \sum_j \frac{(p_j - m_j)^2}{v_j}
\end{equation}
where $f(t_i)$ is the flux at date $t_i$ and $\sigma(t_i)$ is the flux uncertainty at the same date.

SNe Ia form a fairly homogeneous class of objects. Significant departures from typical values of color and stretch could therefore be avoided by setting Gaussian priors on these two parameters: mean $m_c = 0.04$, standard deviation $\sigma_c = 0.15$ for color and $m_{x_1}=0$, $\sigma_{x_1}=1$ for stretch, in agreement with the data. 

\begin{figure}[htb]
 \vskip -.3cm \includegraphics[width=\columnwidth]{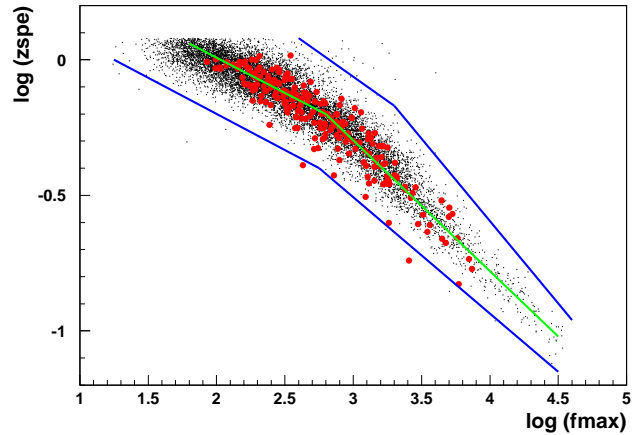}
\caption{Correlation of redshift with the peak flux in $r'$. Black points are for simulated SNe Ia, red dots for the data sample. The blue curves represent the lower and upper bounds set on redshift, and the green curve the mean estimate of redshift for a given peak flux.
}
\label{fig:fmaxvsz}
\end{figure}

Redshift bounds and prior were estimated in the following way. The four band light curves were fitted by the phenomenological form 
\begin{equation}
f(t)\;=\; A \frac{e^{-(t-t_0)/\tau_{\rm fall}}}{1+e^{-(t-t_0)/\tau_{\rm rise}}} \;+ B 
\label{fitformula}
\end{equation}
to obtain estimates of the observed peak luminosities.
As SNe Ia are standard candles, a strong correlation is observed between the redshift and the estimated peak flux $f_{\rm max}$ determined from the fit by equation~\ref{fitformula}.
This correlation, illustrated in figure~\ref{fig:fmaxvsz} for the $r'$ band, allowed setting lower ($z_1$) and upper ($z_2$) bounds on redshift as well as estimating a mean expected redshift value $m_z$, along with an uncertainty $\sigma_z$ taken as the r.m.s. dispersion around $m_z$ for a given observed $f_{\rm max}$.  Typical values for $\sigma_z$ lie between 0.06 and 0.11.
Two SNe Ia lie outside the limits shown in the figure. One has poor sampling (the first point after peak in any of the four bands occurs after 25 days). The other has both a low stretch ($x_1 = -1.2$) and a large intrinsic color ($c=0.36$, close to our limit of 0.4 for our data sample).   
 
Independent estimates of the redshift bounds $z_1$ and $z_2$, mean value $m_z$ and dispersion $\sigma_z$ were computed for the two bands with the best sampling and photometric accuracy, namely $r'$ and $i'$. As a prior on redshift, we used a Gaussian with mean set to the weighted average of the $m_z$ values from $r'$ and $i'$, and with variance taken to be $[\sigma_z(i')^{-2}+\sigma_z(r')^{-2}]^{-1}$. The bounds being quite conservative (over $99.5\%$ of the events contained within the limits), the most restrictive ones were retained to limit the range of redshifts probed. The bounds are only useful to reduce the time consumption of the fit and do not lead to further improvement beyond what is already achieved with the prior on redshift. 

\subsection{Fitting procedure}\label{sec:proc}
In the standard use of SALT2, the fit of the light curves is done at known redshift, and it is best to use only the bands which fall in the rest-frame wavelength range between 290 and 700~nm. In the case studied here, as the redshift is a free parameter, all four bands $g'$, $r'$, $i'$ and $z'$ are used at all times. The priors on color, stretch and redshift defined above were used throughout. 

The first step of the fit consisted in computing the reduced $\chi^2$  for successive values of the redshift separated by $\delta z = 0.1$, within the bounds defined as explained above. To reduce the time consumption, this step is done at fixed $c=m_c$ and $x_1=m_{x_1}$. The value of the redshift $z_{\rm min}|_{c,x_1}$ yielding the minimum reduced $\chi^2$ in this scan was taken as the initial redshift for the second step. In the presence of two minima, both were considered thereafter. Note that with the bounds set on the redshift range, multiple minima only occurred in 3.5\% of the events.  

A finer tuning of the initial redshift was then obtained by setting the color and stretch parameters free and performing a new redshift scan around $z_{\rm min}|_{c,x_1}$, with the same step $\delta z$ as above, until a reduced $\chi^2$ minimum was reached. The values of redshift, color and stretch for the fit that yielded the lowest reduced $\chi^2$ were taken as initial values for the full fit.
The latter was performed with the standard code of the light curve fitter SALT2, letting all five parameters free and initialized as described above. When two minima were observed in step 1, both were taken into account. The solution with the smallest reduced $\chi^2$ at the end of the full fit was chosen. Note that in 71\% of the cases with two $\chi^2$ local minima, both minima ended up with the same set of values for all five parameters (within statistical errors). Only 1\% of the events thus truly needed to go through all 3 steps for both $z_{\rm min}|_{c,x_1}$ to determine the best redshift.
 
 \section{Statistical accuracy of SN Ia photometric redshifts}\label{sec:accuracy}

\begin{table*}[htb]
\begin{center}
\begin{tabular}{lccccccc}
 \cline{1-8}
&  \multicolumn{2}{c}{Simulation} & \multicolumn{5}{c}{Data} \\
Performance& & & Spec. Ia& All Ia& Spec. Ia& All Ia& All Ia\\
criteria& $\zpho^a$ & $\zpho^b$ & $\zpho^a$& $\zpho^a$& $\zpho^b$ & $\zpho^b$ & $z_{\rm gal}$  \\
& \tiny{(1)} & \tiny{(2)} &  \tiny{(3)} &  \tiny{(4)} &  \tiny{(5)} &  \tiny{(6)} &  \tiny{(7)}  \\
\hline
$\sigma_{\Delta z/(1+z)}$ for $z<0.45$ &0.006 & 0.006 &0.018&0.019&0.014 & 0.016& 0.031 \\
$\sigma_{\Delta z/(1+z)}$ for $z>0.45$ &0.029 & 0.027 &0.024&0.027& 0.020 &0.025 & 0.037 \\
Outlier rate $\eta$ & 2.5\% & 1.0\%&2.0\% &3.1\%& 0.4\% & 1.4\% & 5.5\% \\
\hline
\end{tabular}
\caption[]{ Accuracy $\sigma_{\Delta z/(1+z)}$ of the supernova photometric redshift $\zpho$: columns 1 and 2 are for simulated events, columns 3 to 6 for data. For comparison, column 7 indicates the accuracy of the photometric redshift of the host galaxy $\zgal$. The data sample contains 57 objects at $\zspe<0.45$ of which 50 are spectroscopically confirmed, and 232 at $\zspe>0.45$ including 191 spectroscopically confirmed SNe Ia. Superscript $b$ if bounds and prior on redshift used (as explained in section~\ref{sec:procedure}), $a$ otherwise.}
\label{tab:zpho}
\end{center}
\end{table*}

The performance of the method described above was assessed from the two samples of simulated and observed SN Ia light curves. The photometric redshift was compared with the generated one ($z_{\rm gen}$) for simulated events and with the spectroscopic one ($z_{\rm spe}$) for data.

\begin{figure}[htb]
 \vskip -.3cm  \includegraphics[width=\columnwidth]{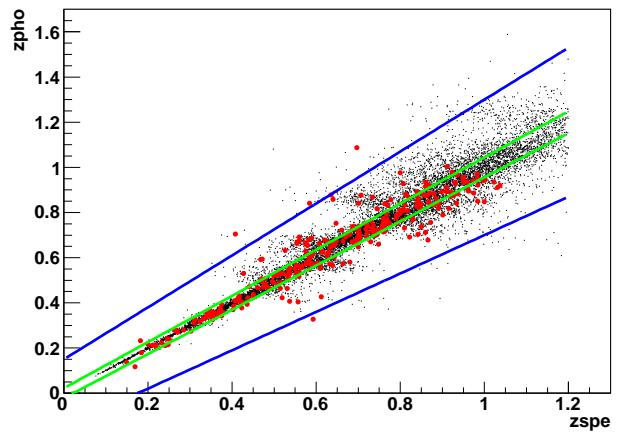}
\caption{
Photometric vs. spectroscopic redshift for the method of this paper. The green lines are for $\zpho = \zspe \pm 0.022(1+\zspe)$ representing the average precision over the entire redshift range, and the blue lines for $\zpho = \zspe \pm 0.15(1+\zspe)$ to visualize the catastrophic redshifts. Red circles are data SNe Ia for which a spectroscopic redshift is available, black points are the simulated SNe Ia.
}
\label{fig:zpho}
\end{figure}

In both cases, 99\% of the events lie within $0.15 \times (1+z)$  of the actual redshift. With the precision defined  as in \cite{bib:coupon2008} by $\sigma_{\Delta z/(1+z)}  \equiv 1.48\times{\rm median}[|\Delta z|/(1+z)]$ (a robust estimate of the r.m.s), and the outlier rate $\eta$,  or rate of catastrophic errors, defined as the proportion of events with $|\Delta z|/(1+z)>0.15$, we obtain a precision of 0.022 and $\eta = 1.4\%$ on average over the entire redshift range\footnote{SALT2 was trained with 40\% of the data used here. Restricting to the complementary set yields results in agreement with the quoted performances within statistical uncertainties.} (see figure~\ref{fig:zpho}). 10\% of the events have $|\Delta z|/(1+z) > 0.07$. For 70\% of these, the reduced $\chi^2$ is smaller than for the fit initialized at the actual redshift.\footnote{Note that the $\chi^2$ comparison is done on the sole contribution of the data points (i.e. not including the contribution of the priors), as there is no redshift prior for the fit initialized at the actual redshift.} The photometric redshift therefore appears, wrongly, to be a better solution than the actual redshift. This is a limitation of any method based on $\chi^2$ minimization. 

The accuracy achieved for two redshift intervals (below and above 0.45) is given in table~\ref{tab:zpho} . The reduced accuracy at high redshift comes from the vanishing SN flux in the $g'$ then $r'$ bands of SNLS, thus enhancing fit degeneracies (see discussion). The redshift dependence of the accuracy is well reproduced by an increase of the spread in $\Delta z$ proportionally to $(1+z)^2$, with $\sigma_{\Delta z/(1+z)^2} = 0.014$ independently of redshift. 

Comparison of columns 2 and 6 shows that $\eta$ for simulation and data are in agreement, given the  uncertainty of 0.7\% on $\eta$ for data. Resolutions at high redshift are also in agreement. At low redshift, it is slightly worse for data, probably due to sources of SN Ia heterogeneity not reproduced in the simulation. 

As visible in columns 5 and 6, the photometric determination of the supernova redshift has very few outliers for spectroscopically confirmed SNe Ia. Most of the catastrophic failures occur for events that have been added by the photometric selection. The increase in $\eta$ between columns 5 and 6 is in agreement with the estimated contamination fraction of 1\% of the total sample. The precision also tends to be slightly better for spectroscopically confirmed events.

An alternative to supernovae photometric redshifts is to use photometric redshifts of the host galaxies, $z_{\rm gal}$. \cite{bib:ilbert2006} have published photometric redshifts for 522286 objects in the four deep fields of SNLS. Only 83\% of potential SNe Ia have a matched host\footnote{The match to a host galaxy was considered successful out to a supernova-host distance of $5\,r_{\rm gal}$ with $r_{\rm gal}$ defined as the half-width of the galaxy in the direction of the event. 9\% of the SNe lie in regions masked in the galaxy catalog because of neighboring bright stars, and 9\% of the remaining SNe are associated to a galaxy lacking a secure $z_{\rm gal}$ for diverse photometric reasons.} with a $z_{\rm gal}$, while SN Ia photometric redshifts can be computed for all objects.
The two photometric redshift determinations are compared in the last two columns of table~\ref{tab:zpho}. Supernova photometric redshifts have a much better precision. 
However, when combined with a photometric selection of SNe Ia, it may lead to samples more contaminated by core-collapse supernovae for instance than when using galactic photometric redshifts which are independent of the light curves. Future projects might want to combine the two techniques to obtain a selection of SNe Ia with reduced contamination from non-Ia supernovae (using fits with redshift fixed at $z_{\rm gal}$ to obtain the selected sample of SNe Ia) while having the redshift accuracy of supernovae photometric redshifts  (refitting the redshift of the selected events to derive $z_{\rm pho}$).

\section{Discussion}

As illustrated in columns 1 and 2 (for the simulation) or 4 and 6 (for the data) of table~\ref{tab:zpho}, the prior set on redshift reduces the rate of catastrophic errors by $\sim 1.5\%$. Their impact on precision, however, is small, because the precision attained is significantly better than the width $\sigma_z$ of the prior. 

We have also checked the impact of assuming that SNe Ia are ``normal'' by removing all the priors on color, stretch and redshift but maintaining the same fitting procedure, i.e. two consecutive $\chi^2$ scans to initialize the values of the parameters for the full fit. The precision obtained on average over the entire redshift range is then of 0.029 and the rate of outliers is significantly increased to $\eta = 9.6\%$. Most of these accidental fits end with extreme colors ($c>2$) assigned to the events. The assumption that SNe Ia are normal therefore improves slightly the accuracy (by $\sim 25\%$) but is crucial to reach a small rate of accidental redshifts (reduction by factor $\sim 10$). 

As inaccurate redshifts were correlated with inaccurate colors, we tried to replace the statistical prior on color by an estimate of the color of each event based on its four-band photometry. This led to little improvement due to the large dispersion around this estimate (often larger than $\sigma_c =0.15$ observed statistically on the sample of data SNe Ia). We thus remained with the prior $m_c=0.04$ and $\sigma_c=0.15$. 

Despite the initialization procedure through two consecutive $\chi^2$ scans, some redshifts are still favored over others. To understand the origin of this problem, two tests were performed. 

\begin{figure}[p]
 \vskip -.7cm
\includegraphics[width=\columnwidth]{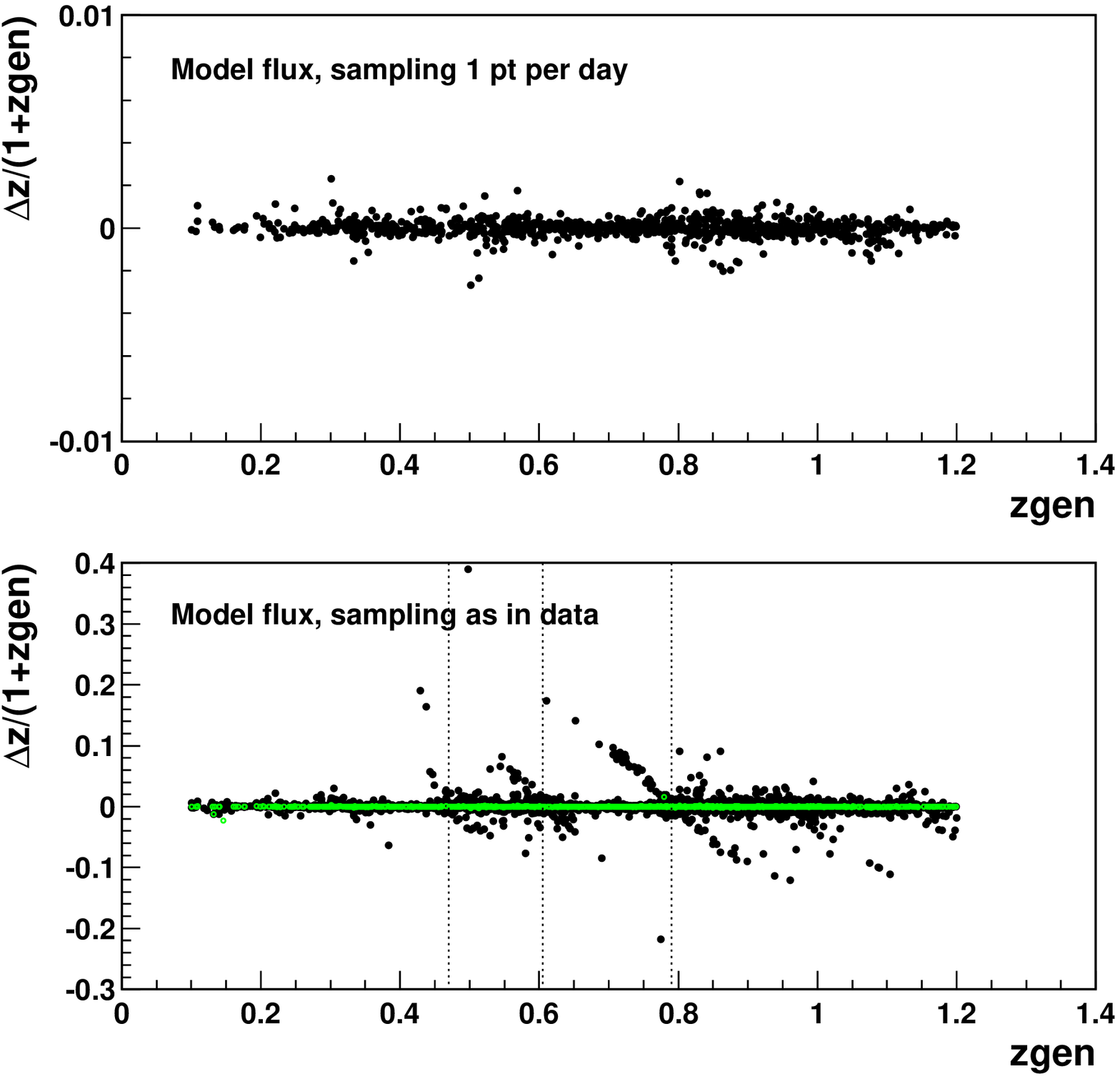}
\caption{
Photometric vs. generated redshift  for simulated supernovae with perfect photometry.
Time sampling is one point per day (top plot) or as in SNLS data (bottom plot, with a different scale). Lines indicate redshifts which tend to be favored by the fit.  Results from sampling  with one point per day superimposed on bottom plot as open green circles.
}
\label{fig:modelFzphozgen}
\vskip -.3cm
\includegraphics[width=\columnwidth]{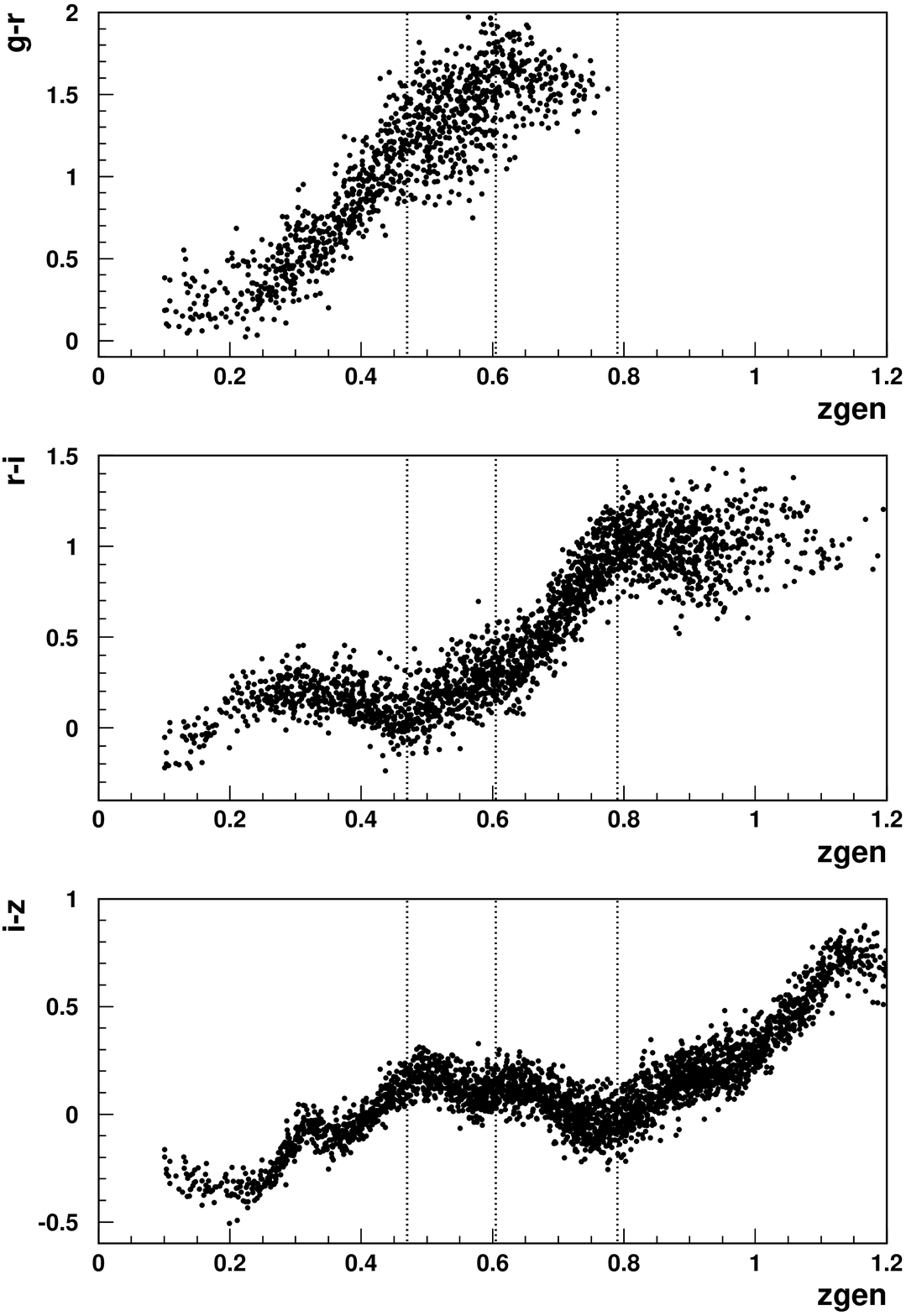}
\caption{
Generated colors as a function of generated redshift. Shown in each plot are the events with a magnitude brighter than 25 in the two concerned filters. Lines indicate redshifts which tend to be favored by the fit.}
\label{fig:modelF3}
\end{figure}

The first one consisted in applying the procedure to perfect simulated light curves with a sampling of one point per night and no flux dispersion. The redshift was determined with an accuracy $\sigma_{\Delta z /(1+z)}<0.001$ over the full redshift range with no redshift favored (figure~\ref{fig:modelFzphozgen}, top plot). 
 
 In the second test, the light curves still had perfect photometry but the sampling was degraded to reproduce that of real data. While the accuracy remained excellent, some accumulations (bottom plot of figure~\ref{fig:modelFzphozgen}) appeared at the same values of $\zpho$ as with the full simulation of figure~\ref{fig:zpho}.  The $z\sim 0.47$ and $z\sim 0.80$ accumulations correspond to turn-overs in the $r-i$ vs. redshift diagram (cf.  figure~\ref{fig:modelF3}). The smaller one at $z\sim 0.60$ corresponds to a short plateau. Since SALT2 fits light curve colors, and as the $r'$ and $i'$ bands are the ones which contribute the most (best sampling and S/N), double redshift values for a given $r-i$ color can indeed lead to the observed degeneracies. The first two can partially be broken with data in the $g'$ band since the $g-r$ vs. redshift diagram is monotonic (this also explains why the turn-over at $z\sim 0.3$ does not lead to redshift accumulations). This is not the case for $z\sim 0.8$ where there is no longer any significant flux in the $g'$ band. 

Note that when reaching perfect photometry, the data $\chi^2$ vanishes for the generated values of the fitted parameters and the priors then play a major role. In the case of redshift degeneracies, the fit converges to the redshift (out of the two possible ones) that minimizes the contribution of the color and stretch priors. Thus, near $\zpho \sim 0.8$, events with large/small intrinsic colors tend to be fitted with over-/under-estimated redshifts. With more data points in a third band (i.e. other than $r'$ and $i'$), the contribution of the priors to the total $\chi^2$ becomes negligible and the fit is truly constrained by the data only. As shown in the top plot of figure~\ref{fig:modelFzphozgen}, a sampling of 1 point every day in all four bands fully breaks the degeneracies. 
 
\begin{figure}[htb]
 \vskip -.7cm \includegraphics[width=\columnwidth]{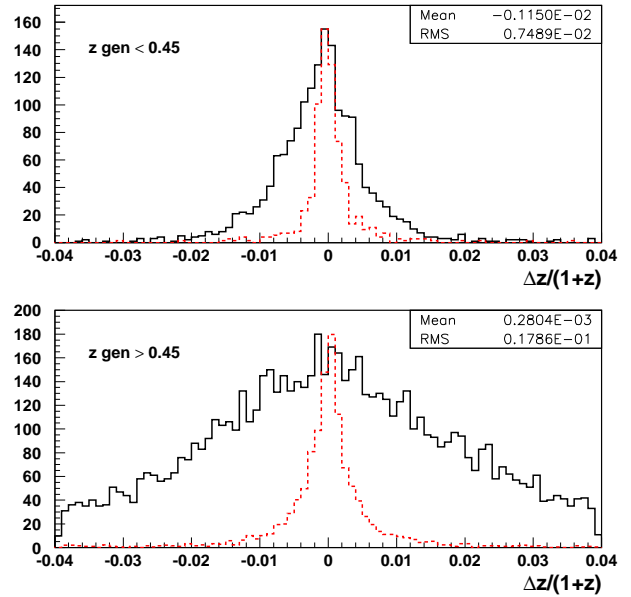}
\caption{
Accuracy of  the redshift determination in the full simulation (solid curve) and in the simulation with perfect photometry but time sampling as in the data (dashed curve). Mean and rms are given for the full simulation. }
\label{fig:modelF}
\end{figure}

As illustrated in figure~\ref{fig:modelF}, the main impact of the photometric uncertainties is to reduce the accuracy with which the redshift is determined by increasing the width of the $\zpho \sim \zspe$ band. Note that the distributions are non-Gaussian, which justifies the use of a median-based estimate of the r.m.s (section~\ref{sec:accuracy}). We have verified that the region thus defined does include at least  $68\%$ of the simulated events whatever the redshift range up to $z\sim 1.1$.

SALT2 requires no assumption on cosmology. A dependence on cosmology could only come from the bounds and prior on redshift. This was tested by simulating SNe Ia in a flat $\Omega_M =1$ Universe where the SNe Ia are more luminous by $\sim 0.5$~mag at $z=0.6$. The bounds are sufficiently loose to be valid even in that extreme case. The prior on redshift is systematically underestimated by $\sim 0.1$, but at the end of the fitting procedure we obtain no bias and the same precision and outlier rate as in table~\ref{tab:zpho} column 2. The only difference is that outliers are more often under- than over-estimates of the redshift. 

\section{Uncertainties and biases}

Although the photometric redshift shows negligible bias on average ($\langle \zpho - \zspe\rangle =0.003$), there is a clear correlation between color and redshift biases: overestimated redshifts are systematically correlated with underestimated colors. 

\begin{figure}[htb]
 \vskip -.3cm \includegraphics[width=\columnwidth]{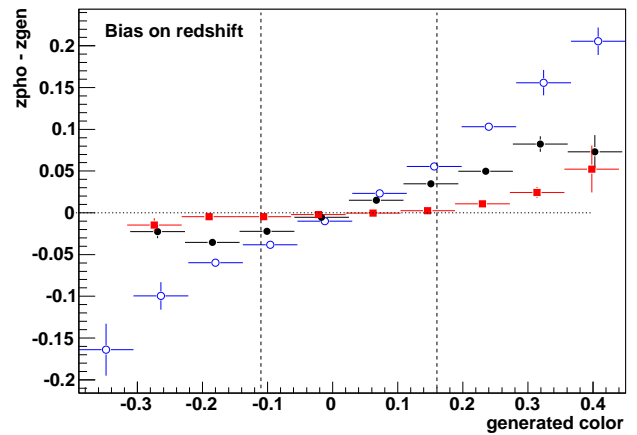}
\caption{
Redshift bias as a function of the supernova intrinsic color for three redshift bins: red squares for $z<0.45$, filled black circles for $0.45<z<0.65$ and open blue circles for $z>0.65$. The vertical lines delimit the 1$\sigma$ color range.}
\label{fig:cbias}
\end{figure}

Supernovae with $z<0.6$ have non-zero flux in at least 3 bands and usually end up with good estimates of both redshift and color. As the redshift increases however, the loss of information (due to unmeasurably small fluxes in the bluest bands) gives increasingly more weight to the color prior compared to the data. The fitted color thus tends to converge to that of the prior ($m_c$), leading to a color-dependent bias (cf figure~\ref{fig:cbias}). This is particularly clear near redshift degeneracies (cf figure~\ref{fig:zpho}), where events with positive (resp. negative) intrinsic colors populate the $\zpho > z_{\rm gen}$ (resp. $\zpho < z_{\rm gen}$) accumulations. 

To reduce this bias and improve the color and redshift determination for ``colored'' SNe ($|c_{\rm gen}-m_c| > 0.1$ where $c_{\rm gen}$ is the generated color) in SNLS, we performed an additional SALT2 fit as follows : the mean and $\sigma$ of the Gaussian prior on $x_1$ are set to the best fit value of $x_1$ and its uncertainty from the procedure of section~\ref{sec:procedure}, a Gaussian prior on redshift is used with mean set to the value found previously and $\sigma$ taken as $0.014\times (1+\zpho)^2$. With these stronger constraints, we eliminate the prior on color. 

\begin{figure}[htb]
 \vskip -.7cm \includegraphics[width=\columnwidth]{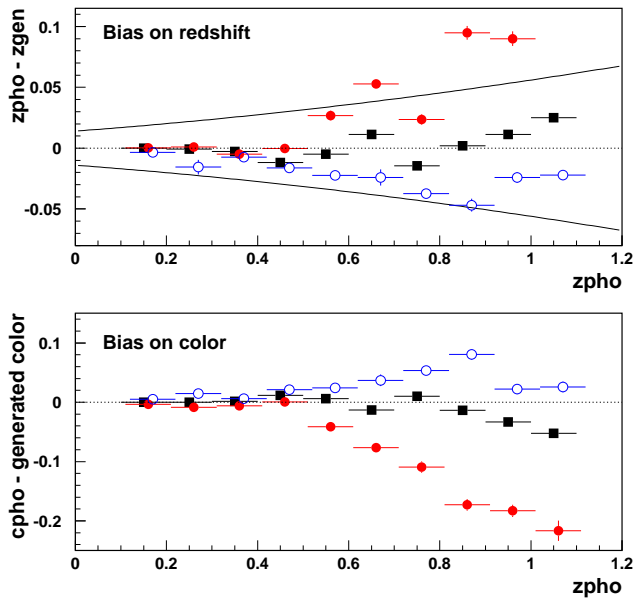}
\caption{
Top: bias on redshift as a function of the fitted redshift with the additional SALT2 fit  for three bins of intrinsic (generated) color $c_{\rm gen}$: black squares for $|c_{\rm gen}-m_c|<0.1$, filled red circles for $c_{\rm gen}-m_c>0.1 $ and open blue circles for $c_{\rm gen}-m_c<-0.1 $. The curves illustrate the $0.014(1+z)^2$ statistical uncertainty. Bottom: bias on color as a function of the fitted redshift. Same color code as above.}
\label{fig:bias}
\end{figure}

Above a redshift of 0.65, the additional fit reduces the bias $z_{\rm pho} - z_{\rm gen}$ by $\sim 0.02$ for blue SNe ($c_{\rm gen}-m_c < -0.1$) while yielding a smaller but systematic improvement on red ($c_{\rm gen}-m_c > 0.1$) SNe. The color determination is also systematically improved: above a redshift of 0.65, the color bias is reduced by 0.04 for blue SNe and by 0.01 for red ones.
Fits to events with $c_{\rm gen}\sim m_c$ (black squares in figure~\ref{fig:bias}), or to  low redshift events where the data have more weight than the priors, are unchanged: these events were already well fitted. 

Figure~\ref{fig:bias} illustrates the remaining bias on fitted redshift or fitted color as a function of redshift, for several ranges of intrinsic color. At high redshift, the reddest (and thus also the faintest) SNe exhibit the largest bias, emphasizing the lack of another filter (beyond the $z$ band) with measurable flux to determine the SN parameters. On average for events passing the preselection filter, however, the bias remains small, with $\zpho-\zgen < 0.027$ whatever the redshift up to a $z \sim 1.1$, and $\langle \zpho-\zgen \rangle = 0.008$ on average over the full redshift range between 0 and 1.1. The statistical uncertainty derived in section~\ref{sec:accuracy} is indicated in figure~\ref{fig:bias}. It encompasses about 68\% of the events at all redshifts. 

\section{Conclusions}

The study presented in this paper shows that SN Ia photometric redshifts can be obtained with an accuracy exceeding that of current host galaxy photometric redshifts. 
The method was applied to simulated light curves as well as to a sample of SNe Ia from the first three years of SNLS data for which the redshift was known from spectroscopy. The accuracy reaches $\sigma_{\Delta z/(1+z)} = 0.022$ on average over the entire redshift range, with $\sigma_{\Delta z/(1+z)} = 0.016$ for $z<0.45$ and $\sigma_{\Delta z/(1+z)} = 0.025$ for $0.45<z<1.1$. Degeneracies in the fit above $z \sim 0.45$,  due to vanishing flux in the bluest band of SNLS, are responsible for the lower accuracy at high redshifts. 
The spread in $\Delta z /(1+z)^2$ is seen to be independent of redshift, with $ \sigma_{\Delta z/(1+z)^2} = 0.014$. 

Red (resp. blue) SNe tend to have an overestimated (resp. underestimated) redshift. The net bias however on the fitted redshift, after integration over the SN population, remains smaller than 0.027 whatever the redshift up to $z \sim 1.1$, with a net bias $\langle \zpho-\zgen \rangle$ of 0.008 on average.

For only 1.4\% of the events do we obtain a catastrophic error ($|\Delta z|/(1+z)>0.15$) on the estimate of the redshift, and for 10\% an error larger than 0.07. The rate of catastrophic failures is reduced to 0.4\% for spectroscopically confirmed SNe Ia.

The method is illustrated with SNLS data but can be adapted to other supernova samples. 

\begin{acknowledgements}
SNLS relies on observations with MegaCam, a joint project of
CFHT and CEA/DAPNIA, at the Canada-France-Hawaii Telescope (CFHT)
which is operated by the National Research Council (NRC) of Canada, the
Institut National des Science de l'Univers of the Centre National de la
Recherche Scientifique (CNRS) of France, and the University of Hawaii. This
work is based in part on data products produced at the Canadian
Astronomy Data Centre as part of the Canada-France-Hawaii Telescope Legacy
Survey, a collaborative project of the National Research Council of
Canada and the French Centre national de la recherche scientifique.
SNLS also relies on observations obtained at the European
Southern Observatory using the Very Large Telescope on
the Cerro Paranal (ESO Large Programme 171.A-0486), 
on observations (programs GN-2006B-Q-10, GN-2006A-Q-7, GN-2005B-Q-7, GS-2005B-Q-6, GN-2005A-Q-11, GS-2005A-Q-11, GN-2004B-Q-16, GS-2004B-Q-31, GN-2004A-Q-19, GS-2004A-Q-11, GN-2003B-Q-9, and GS-2003B-Q-8) obtained at the
Gemini Observatory, which is operated by the Association
of Universities for Research in Astronomy, Inc., under a
cooperative agreement with the NSF on behalf of the Gemini
partnership: the National Science Foundation (United States),
the Particle Physics and Astronomy Research Council (United
Kingdom), the National Research Council (Canada), CONICYT
(Chile), the Australian Research Council (Australia), CNPq
(Brazil) and CONICET (Argentina), and on observations
obtained at the W.M. Keck Observatory, which is operated
as a scientific partnership among the California Institute of
Technology, the University of California and the National
Aeronautics and Space Administration. The Observatory was
made possible by the generous financial support of the W.M.
Keck Foundation.
MS acknowledges support from the Royal Society.

\end{acknowledgements}

\end{document}